\documentclass[preprint,12pt]{elsarticle}

\usepackage{graphicx}
\usepackage{amssymb}
\usepackage{amsmath}
\newtheorem{theorem}{Theorem}

\newtheorem{proposition}{Proposition}
\newtheorem{corollary}{Corollary}

\usepackage{lineno}

\journal{Journal Name}

\begin{document}

\begin{frontmatter}


\title{On Domination Coloring in Graphs}

 \author{Yangyang Zhou}
 \ead{zyy_eecs@pku.edu.cn}
 \author{Dongyang Zhao}
 \address{Peking University, Beijing, China}


\begin{abstract}
A domination coloring of a graph $G$ is a proper vertex coloring of $G$ such that each vertex of $G$ dominates at least one color class, and each color class is dominated by at least one vertex. The minimum number of colors among all domination colorings is called the domination chromatic number, denoted by $\chi_{dd}(G)$. In this paper, we study the complexity of this problem by proving its NP-Completeness for arbitrary graphs, and give general bounds and characterizations on several classes of graphs. We also show the relation between dominator chromatic number $\chi_{d}(G)$, dominated
chromatic number $\chi_{dom}(G)$, chromatic number $\chi(G)$, and domination number $\gamma(G)$. We present several results on graphs with $\chi_{dd}(G)=\chi(G)$. 
\end{abstract}

\begin{keyword}
Coloring \sep Domination \sep Domination coloring \sep Domination chromatic number


\end{keyword}

\end{frontmatter}


\section{Introduction}
\label{S:1}

Graphs considered in this paper are finite, simple and undirected. We start with some basic notions and definitions of graphs. Let $G=(V, E)$ be a graph with $n=|V|$ and $m=|E|$. For any vertex $v \in V(G)$, the open neighborhood of $v$ is the set $N(v)=\{u | uv \in E(G)\}$ and the closed neighborhood is the set $N[v]=N(v) \cup \{v\}$. Similarly, the open and closed neighborhoods of a set $X \subseteq V$ are respectively $N(X)=\bigcup_{x \in X}N(x)$ and $N[X]=N(X) \cup X$. The degree of a vertex $v \in V$, denoted by $deg(v)$, is the cardinality of its open neighborhood. The maximum and minimum degree of a graph $G$ is denoted by $\Delta(G)$ and $\delta(G)$, respectively. Given a set $X \subseteq V$, we denote by $G[X]$ the subgraph of $G$ induced by $X$. Given any graph $H$, a graph $G$ is $H$-free if it does not have any induced subgraph isomorphic to $H$. We denote by $P_n$ the path on $n$ vertices and by $C_n$ the cycle on $n$ vertices. A tree is a connected acyclic graph. The complete graph on $n$ vertices is denoted by $K_n$ and the complete graph of order 3 is called a triangle. The complete bipartite graph with classes of orders $r$ and $s$ is denoted by $K_{r,s}$. A star is a graph $K_{1,k}$ with $k \geq 1$. A bi-star is a graph formed by two stars by adding an edge between the center vertices. For any undefined terms, the reader is referred to the book of Bondy and Murty \cite{Bondy:2008}.

A proper vertex $k$-coloring of a graph $G = (V, E)$ is a mapping $f : V \rightarrow \{1,2,\cdots,k\}$ such that any two adjacent vertices receive different colors. In fact, this problem is equivalent to the problem of partitioning the vertex set of $G$ into $k$ independent sets $\{V_1,V_2,\cdots,V_k\}$ where $V_i = \{x \in V | f(x) = i\}$. The set of all vertices colored with the same color is called a color class. The chromatic number of $G$, denoted by $\chi(G)$, is the minimum number of colors among all proper colorings of $G$. A graph admitting a proper $k$-coloring is said to be $k$-colorable, and it is said to be $k$-chromatic if its chromatic number is exactly $k$. 

A dominating set $S$ is a subset of the vertices in a graph $G$ such that every vertex in $G$ either belongs to $S$ or has a neighbor in $S$. The domination number $\gamma(G)$ is the minimum cardinality of a dominating set of $G$. A $\gamma(G)$-set is a dominating set of $G$ with minimum cardinality.

Graph coloring and domination are two major areas in graph theory and both have been well studied. There exist plenty of variants of classical graph coloring \cite{Chen:1998, Malaguti:2010}. Also, excellent surveys on the fundamentals of domination and several advanced topics are given in \cite{Haynes:199801} and \cite{Haynes:199802}, respectively. Moreover, graph coloring and domination problems are often in relation. Chellali and Volkmann \cite{Chellali:2004} showed some relations between the chromatic number and some domination parameters in the graph. Indeed, Hedetniemi et al. \cite{Hedetniemi:2009} introduced the concepts of dominator partition of a graph. Motivated by \cite{Hedetniemi:2009}, Gera et al. \cite{Gera:2006} proposed the dominator coloring as a proper coloring such that every vertex has to dominate at least one color class (possibly its own class) in 2006. The minimum number of colors among all dominator colorings of $G$ is the dominator chromatic number of $G$, denoted by $\chi_d(G)$. Gera studied further this coloring problem in \cite{Gera:200701, Gera:200702}. More results on the dominator coloring could be found in \cite{Chellali:2012, Boumediene:2012, ARUMUGAM:2012, Bagan:2017}. In 2015, Boumediene et al. \cite{Boumediene:2015} introduced the dominated coloring as a proper coloring where every color class is dominated by at least one vertex. The minimum number of colors among all dominated colorings of $G$ is the dominated chromatic number of $G$, denoted by $\chi_{dom}(G)$.

For problems mentioned above, the domination property is defined either on vertices or on color classes. Indeed, the color classes in a dominator coloring are not necessarily all dominated by a vertex, and the vertices in a dominated coloring are not necessarily all dominates a color class. In this paper, we introduce the domination coloring that both of the vertices and color classes should satisfy the domination property. A domination coloring of a graph $G$ is a proper vertex coloring of $G$ such that each vertex of $G$ dominates at least one color class, and each color class is dominated by at least one vertex. The minimum number of colors among all domination colorings, denoted by $\chi_{dd}(G)$, is called the domination chromatic number.

We concern with connected graphs only. The aim of this paper is to study properties and realizations of the dominator chromatic number. In Section 2, we analyse the basic complexity of the domination coloring parblem. In Section 3, we present the domination chromatic number for classes of graphs, and find bounds and characterization results. We investigate some realization results in Section 4, and pose open questions in Section 5. 

\section{Basic complexity results}

This section focuses on the complexity study of the domination coloring problem. Whether an arbitrary graph admits a domination coloring with at most $k$ colors? We aim to this decision problem and give the following formalization:

\begin{itemize}
\item $k$-Domination Coloring Problem:\\
Instance: A graph $G=(V, E)$ without isolated vertices and a positive integer $k$.\\
Question: Is there a domination coloring of $G$ with at most $k$ colors?
\end{itemize}

\begin{theorem}
\label{th1}
For $k \geq 4$, the $k$-domination coloring problem is NP-Complete.
\end{theorem}

\begin{figure}[h]
\centering\includegraphics[width=0.6\linewidth]{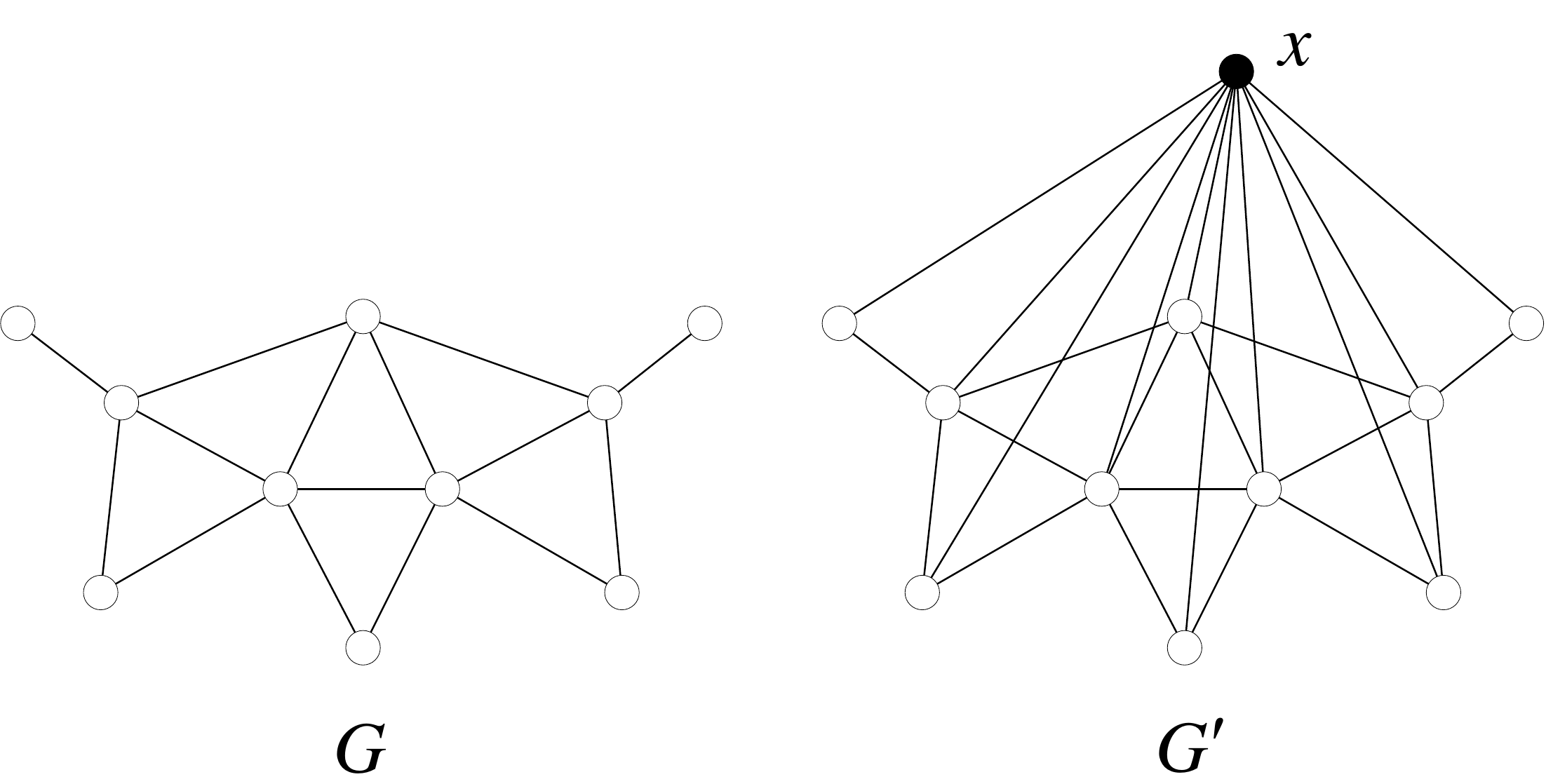}
\caption{The graphs $G$ and $G'$}
\label{f1}
\end{figure}

Proof. The $k$-Domination Coloring Problem is in NP, since verifying if a coloring is a domination coloring could be performed in polynomial time. Now, we give a polynomial time reduction from the $k$-Coloring Problem which is known to be NP-Complete, for $k \geq 3$. Let $G=(V,E)$ be a graph without isolated vertices. We construct a graph $G'$ from $G$ by adding a new vertex $x$ to $G$ and adding edges between $x$ and every vertex of $G$. That is, $x$ is a dominating vertex of $G'$, as shown in Figure \ref{f1}. We show that $G$ admits a proper coloring with $k$ colors if and only if $G'$ admits a domination coloring with $k + 1$ colors. 
\hfill $\square$

First, we prove the necessity. Let $f$ be a proper $k$-coloring of $G$, and the corresponding color classes set is $\{V_1, V_2, \cdots, V_k\}$. We construct a $(k+1)$-domination coloring $f'$ of $G'$ with the color classes set $\{V_{1}^{'}=V_{1}, V_{2}^{'}=V_{2}, \cdots, V_{k}^{'}=V_{k}, V_{k+1}^{'}=\{x\}\}$. It is easy to see that $f'$ is a domination coloring of $G'$ since

\begin{enumerate}
\item $f'$ is proper;
\item Each vertex other than $x$ dominate at least the color class containing $x$ and $x$ dominate all color classes of $f'$; 
\item Each color class other than $\{x\}$ is dominated by $x$ and the color class containing $x$ is dominated by any other vertex.
\end{enumerate}

Then, we prove the sufficiency. Let $f'$ be a $(k+1)$- domination coloring of $G'$, and $\{V_{1}^{'}, V_{2}^{'}, \cdots, V_{k}^{'}, V_{k+1}^{'}\}$ is the color classes set. Since $f'$ is proper, there exists a color class $V_{i}^{'}$ such that $V_{i}^{'}=\{x\}$. So, we can construct a proper $k$-coloring of $G$ by removing the color class $V_{i}^{'}$ from $f'$.

From the above, the $k$-Domination Coloring Problem is NP-Complete, for $k \geq 4$.

\section{Domination coloring for classes of graphs}
 
In this section, we study some properties of the domination coloring and basic results on typical classes of graphs.

Let $G$ be a connected graph with order $n \geq 2$. Then at least two different colors are needed in a domination coloring since there are at least two vertices in $G$ adjacent to each other. Moreover, if each vertex receives a unique color, then both the vertices and color classes satisfy the domination property. Clearly, we get a domination coloring of $G$ with $n$ colors. Thus,

\begin{equation}
\label{d:d}
2 \leq \chi_{dd}(G) \leq n
\end{equation}

Gera et al. \cite{Gera:2006} introduced the Inequalities (\ref{d:tor}) for the dominator chromatic number $\chi_{d}(G)$ and Boumediene et al. \cite{Boumediene:2015} obtained Inequalities (\ref{d:ted}) for the dominated chromatic number $\chi_{dom}(G)$. Also, we can get a similar inequality for the domination chromatic number $\chi_{dd}(G)$. 

\begin{equation}
\label{d:tor}
max\{\chi(G),\gamma(G)\} \leq \chi_d(G) \leq \chi(G)+\gamma(G)
\end{equation}

\begin{equation}
\label{d:ted}
max\{\chi(G),\gamma(G)\} \leq \chi_{dom}(G) \leq \chi(G) \cdot \gamma(G)
\end{equation}

\begin{proposition}
\label{pro1}
Let $G$ be a graph without isolated vertices, then
$$max\{\chi(G),\gamma(G)\} \leq max\{\chi_d(G),\chi_{dom}(G)\} \leq \chi_{dd}(G) \leq \chi(G) \cdot \gamma(G).$$
\end{proposition}

Proof. It is easy to check the left two parts of the inequality. Consider a graph $G$ and a $\gamma(G)$-set $D$ of $G$. We obtain a domination coloring of $G$ by giving distinct colors to each vertex $x$ of $D$ and at most $\chi(G)-1$ new colors to the vertices of $N(x)$. Hence, we totally use at most $\gamma(G)+(\chi(G)-1) \cdot  \gamma(G) = \chi(G) \cdot \gamma(G)$ colors. So, $\chi_{dd}(G) \leq \chi(G) \cdot \gamma(G)$.
\hfill $\square$

The bound of Proposition \ref{pro1} is tight for complete graphs. Since every planar graph is 4-colorable \cite{Appel:1976}, the following is straightforward:

\begin{corollary}
Let $G$ be a planar graph without isolated vertices, then $\chi_{dd}(G) \leq 4 \gamma(G)$.
\end{corollary}

\begin{theorem}
\label{th2}

(1) For the path $P_n$, $n \geq 2$, 
$$\chi_{dd}(P_n)=2 \cdot \lfloor\frac{n}{3}\rfloor + mod(n,3)$$

(2) For the cycle $C_n$,
$$\chi_{dd}(C_n)=
\begin{cases}
2,& \text{n=4}\\
3,& \text{n=3, 5}\\
2 \cdot \lfloor\frac{n}{3}\rfloor + mod(n,3),& \text{otherwise}
\end{cases}$$

(3) For the complete graph $K_n$, $\chi_{dd}(K_n)=n$;

(4) For the complete $k$-partite graph $K_{a_1,a_2,\cdots,a_k}$, $\chi_{dd}(K_{a_1,a_2,\cdots,a_k})=k$;

(5) For the complete bipartite graph $K_{r,s}$, $\chi_{dd}(K_{r,s})=2$;

(6) For the star $K_{1,n}$, $\chi_{dd}(K_{1,n})=2$;

(7) For the wheel $W_{1,n}$,
$$\chi_{dd}(W_{1,n})=
\begin{cases}
3,& \text{n is even}\\
4,& \text{n is odd}
\end{cases}$$
\end{theorem}

Proof. (1) Let $P_n=v_1v_2 \cdots v_n$. By the definition of domination coloring, we discover that at most two non-adjacent vertices are allowed in a color class, if not, there exist no vertex dominating this color class. On the other hand, the vertex adjacent to both of the two vertices of a color class must be the unique vertex of some color class. For convenience, let $P_5=v_1v_2v_3v_4v_5$ be a $P_5$-subgraph of $P_n$. If vertices $v_1$ and $v_3$ are in a color class, then $v_2$ must be the unique vertex of a color class. If not, $v_2$ and $v_4$ are partitioned in a color class, which will result in $v_4$ cannot dominate any color class. So, every three vertices of $P_n$ need be partitioned in two color classes, and each of the rest vertices forms its own color class. Clearly, it is an optimal domination coloring of $P_n$. Thus, $\chi_{dd}(P_n)=2 \cdot \lfloor\frac{n}{3}\rfloor + mod(n,3)$.

(2) For $n=3, 4, 5$, the result follows by inspection. For $n \geq 6$, it is not hard to find the case is similar to the path $P_n$. As the discussion in (1), the result follows.

(3) For the complete graph $K_n$, $\chi (K_n)=n$. By Proposition \ref{pro1} and Inequation (\ref{d:d}),  $\chi_{dd}(K_n)=n$.

(4) Let $K_{a_1,a_2,\cdots,a_k}$ be the complete $k$-partite graph, and $V_i (1 \leq i \leq k)$ be the $k$-partite sets. Then $\chi_{dd}(K_{a_1,a_2,\cdots,a_k}) \geq \chi(K_{a_1,a_2,\cdots,a_k}) = k$. Also, the coloring that assigns color $i$ to each partite set $V_i (1 \leq i \leq k)$ is a domination coloring. The result follows.

(5) and (6) are special cases of (4).

(7) Let $W_{1,n}$ be the wheel with order $n+1$. Since,
$$\chi(W_{1,n})=
\begin{cases}
3,& \text{$n$ is even}\\
4,& \text{$n$ is odd}
\end{cases}$$
and the corresponding proper colorings are also domination colorings, the result follows.
\hfill $\square$

One note on the domination chromatic number is that for a given graph $G$, and a subgraph $H$ of $G$, the domination chromatic number of $H$ can be smaller or larger than the domination chromatic number of $G$. That is to say, induction may be not useful when we want to find the domination chromatic number of a graph. As an example, consider the graph $G = K_n$ and $H = P_2$, then $\chi_{dd}(K_n) = n \geq 2 = \chi_{dd}(P_2)$, and consider the graph $G = K_{n,n}$ and $H = P_{2n}$, then $\chi_{dd}(K_{n,n}) = 2 \leq 2 \cdot \lfloor\frac{2n}{3}\rfloor + mod(2n,3) = \chi_{dd}(P_{2n})$.

\begin{proposition}
\label{pro2}
Let $G$ be a connected graph with order $n$. Then
 $\chi_{dd}(G) = 2$ if and only if $G = K_{r,s}$ for $r, s \in \mathbf{N}$.
\end{proposition}

Proof. By Theorem \ref{th1} (5), if $G=K_{r,s}$, then $\chi_{dd}(G)=2$. We just need to prove the necessity. 

Let $G$ be a connected graph such that $\chi_{dd}=2$, and $V_1$ and $V_2$ are the two color classes. If $|V_1| = 1$ or $|V_2| = 1$, then $G = K_{1,n-1}$. So, suppose that $|V_1| \geq 2$ and $|V_2| \geq 2$. For any vertex $x \in V_1$, since $|V_1| \geq 2$, it follows that $x$ dominates color class $V_2$. Similarly for any vertex in $V_2$. Thus, each vertex of $V_1$ is adjacent to each vertex of $V_2$, and both $V_1$ and $V_2$ are independent. So $G = K_{r,s}$ for $r,s \in \mathbf{N}$, and the result follows.
\hfill $\square$

\begin{proposition}
\label{pro3}
Let $G$ be a connected graph with order $n$. Then
 $\chi_{dd}(G) = n$ if and only if $G = K_n$ for $n \in \mathbf{N}$.
\end{proposition}

Proof. By Theorem \ref{th2} (3), $\chi_{dd}(G)=n$, if $G=K_n$. We only need to prove the necessity. 

Let $G$ be a connected graph with $\chi_{dd}(G)=n$. Suppose that $G \neq K_n$. Thus, there exist at least two vertices, say $x$ and $y$, such that they are not adjacent in $G$. Now, we define a coloring of $G$ in which $x$ and $y$ receive the same color, and each of the remaining vertices receive a unique color. This is a domination coloring, so $\chi_{dd}(G) \leq n-1$, a contradiction. Thus, $G = K_n$, and we obtain the result.
\hfill $\square$

\begin{figure}[h]
\centering\includegraphics[width=0.3\linewidth]{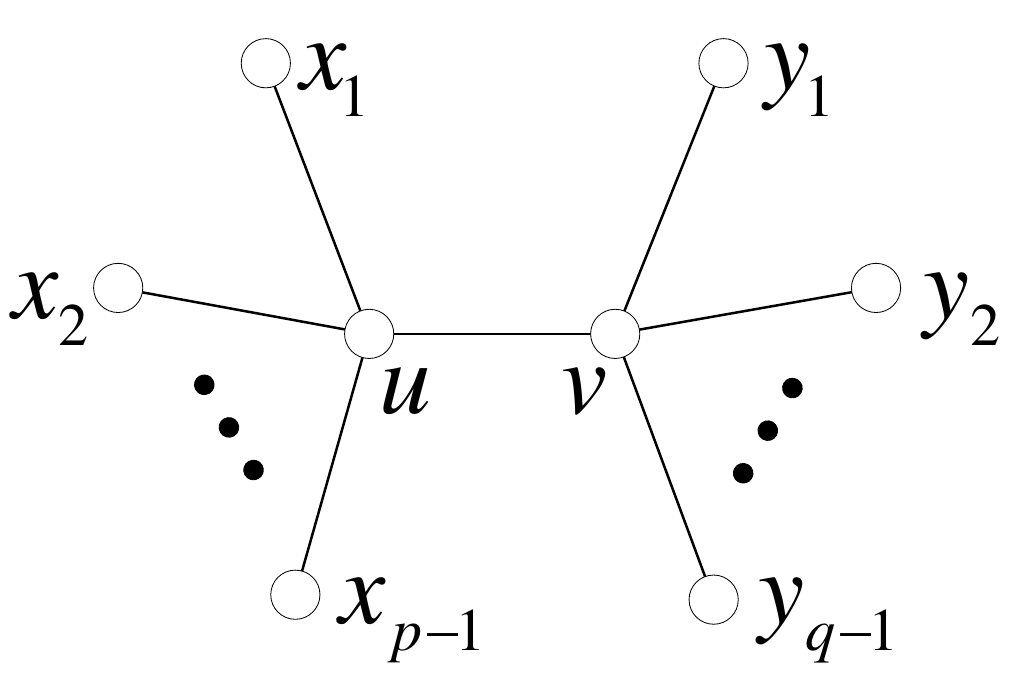}
\caption{The bi-star $S_{p,q}$}
\label{f2}
\end{figure}

Next, we consider the bi-stars. Let $B_{p,q}$ be the bi-star with central vertices $u$ and $v$, where $deg(u) = p \geq 2$ and $deg(v) = q \geq 2$. Let $X=\{x_1,x_2,\cdots,x_{p-1}\}$ and $Y=\{y_1,y_2,\cdots,y_{q-1}\}$. Obviously, $N(u) = X \bigcup \{v\}$ and $N(v) = Y \bigcup \{u\}$, as shown in Figure \ref{f2}.

\begin{theorem}
\label{th3}
For the bi-star $S_{p,q}$ with $p + q \geq 4$, $\chi_{dd}(S_{p,q})=4$.
\end{theorem}

Proof. Consider a proper coloring of $S_{p,q}$ in which the color classes $V_1 = \{u\}$, $V_2 = \{v\}$, $V_3 = X$, and $V_4 = Y$. Then, each vertex in the set $\{u\} \bigcup X$ dominates the color class $V_1$, and each vertex in the set $\{v\} \bigcup Y$ dominates the color class $V_2$. Also, the color class $V_1$ is dominated by any vertex in $V_3$, $V_2$ is dominated by any vertex in $V_4$, $V_3$ is dominated by vertex $u$, and $V_4$ is dominated by vertex $v$. Therefore, this is a domination coloring, and $\chi_{dd}(S_{p,q}) \leq 4$.

By the Lemma 2.2 in \cite{Gera:2006}, $\chi_{d}(S_{p,q})=3$. So, $3 \leq \chi_{dd}(S_{p,q}) \leq 4$. Suppose that $\chi_{dd}(S_{p,q})=3$. It will be result in that each vertex in $X$ or each vertex in $y$ does not dominate a color class. Thus, $\chi_{dd}(S_{p,q})=4$.
\hfill $\square$

\begin{theorem}
\label{th4}
For the Petersen graph $P$, $\chi_{dd}(P)=5$.
\end{theorem}

\begin{figure}[h]
\centering\includegraphics[width=0.3\linewidth]{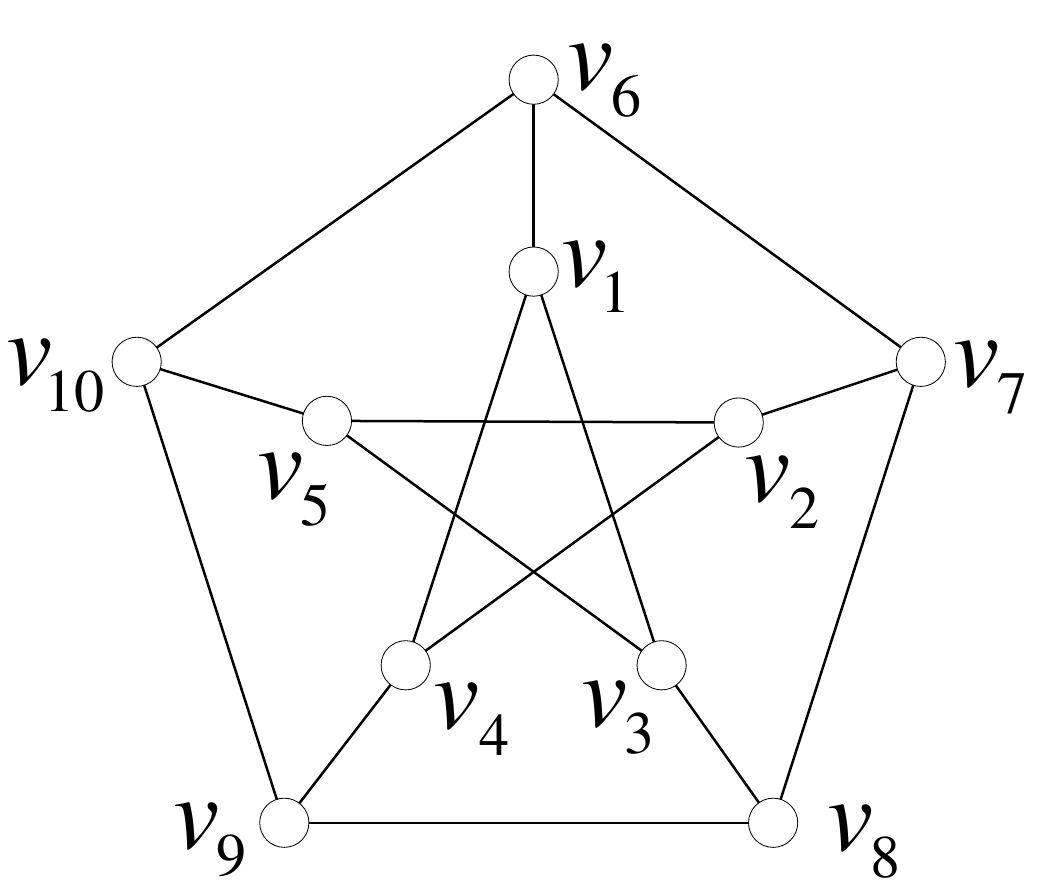}
\caption{The Petersen graph}
\label{f3}
\end{figure}

Proof. It is easy to check $\{\{v_1,v_2,v_9\},\{v_3,v_4,v_6\},\{v_5,v_7\},\{v_8\},\{v_{10}\}\}$ is a domination coloring of the Petersen graph, as shown in Figure \ref{f3}. So, $\chi_{dd}(P) \leq 5$. By Proposition 2.1 in \cite{Bagan:2017}, $\chi_{dom}(P)=4$ and $\chi_{d}(P)=5$. Thus, $\chi_{dd}(P) \geq 5$ by Proposition \ref{pro1}. The result follows.
\hfill $\square$

\begin{proposition}
Let $G$ be a connected graph with order $n$ and maximum degree $\Delta$, then $\chi_{dd}(G) \geq \frac{n}{\Delta}$.
\end{proposition}

Proof. Consider a minimum domination coloring of $G$. Since $G$ is $S_{\Delta + 1}$-free, any color class would not have more than $\Delta$ vertices; otherwise, a vertex dominating such color class will induce a star of order at least $\Delta + 2$, a contradiction. So, $\chi_{dd}(G) \geq \frac{n}{\Delta}$.
\hfill $\square$

\begin{theorem}
\label{th5}
Let $G$ be a connected triangle-free graph, then $\chi_{dd}(G) \leq 2\gamma(G)$.
\end{theorem}

Proof. Consider a minimum dominating set $S$ of $G$. Color every vertex of $S$ with a new color. Since $G$ does not contain any triangle, the set of neighbors of every vertex of $S$ is an independent set. So, a second new color is given for each neighborhood. Obviously, this is a proper coloring of $G$ with $2|S|$ colors, which satisfies that every vertex dominates at least one color class, and every color class is dominated by at least one vertex. Thus, $\chi_{dd}(G) \leq 2\gamma(G)$.
\hfill $\square$

\section{Graphs with $\chi_{dd}(G)=\chi(G)$}

For any graph $G$, we have $\chi_{dd}(G) \geq \chi(G)$. In this section, we investigate graphs for which $\chi_{dd}(G)=\chi(G)$.

\begin{theorem}
\label{th6}
Let $G$ be a connected graph, if $\gamma(G) = 1$, then $\chi_{dd}(G)=\chi(G)$.
\end{theorem}

Proof. Let $G$ be a connected graph of order $n$ with $\gamma(G) = 1$. It then follows that $K_{1,n-1}$ is a spanning subgraph of $G$. Let $v$ be a vertex of degree $n-1$ in $G$. Since $v$ is adjacent to all the vertices of $V(G) \ v$, a minimum coloring of $G$ uses $\chi(G - v) + 1$ colors. Also, this coloring is also a domination coloring of $G$, where each vertex dominates the color class of $v$, the color class $v$ is dominated by any other vertex and each of other color classes is dominated by $v$. So, $\chi_{dd}(G) \leq \chi(G)$, and then $\chi_{dd}(G)=\chi(G)$.
\hfill $\square$

In fact, Theorem \ref{th6} directly follows from Proposition \ref{pro1}.

A unicyclic graph is a graph which contains only one cycle. In the following， we characterize unicyclic graphs with $\chi_{dd}=\chi$.

\begin{theorem}
\label{th7}
Let $G$ be a connected unicyclic graph. Then $\chi_{dd}(G)=\chi(G)$ if and only if $G$ is isomorphic to $C_3$ or $C_4$ or $C_5$ or the graph obtained from $C_3$ by attaching any number of leaves at one vertex of $C_3$.
\end{theorem}

Proof. For the sufficiency, the result is obvious if $G$ is the graph meet conditions. We consider only the necessity. Let $G$ be a connected unicyclic graph with $\chi_{dd}(G)=\chi(G)$, and $C$ the unique cycle of $G$. 

Case 1. If $C$ is an even cycle, then $\chi(G)=2$ and $\chi_{dd}(G)=2$. It follows that $G$ cannot contain any other vertices not on $C$, otherwise $\chi_{dd}(G) \geq 3$. By Theorem \ref{th2}(2), $G = C_4$. 

Case 2. If $C$ is an odd cycle, then $\chi_{dd}(G)=\chi(G)=3$. Suppose there exists a support vertex $x$ not on $C$. Since $x$ or the leaf is a color class in each $\chi_{dd}$-coloring of $G$, it follows that $\chi_{dd}(G) \geq 4$, which is a contradiction. Hence, all the support vertices lie on $C$, and any vertex not on $C$ is a leaf. Morever, the number of support vertices is at most one. Otherwise, it follows that some color class does not be dominated, since there exists some $\chi_{dd}$-coloring of $G$ in which every support vertex appears as a singleton color class. 

Case 2.1. If $|C|=3$, then $G$ is isomorphic to $C_3$ or the graph obtained from $C_3$ by attaching any number of leaves at exactly one vertex of $C_3$.
 
Case 2.2. Suppose that $|C| \geq 5$. If there exists a support vertex $x$ on $C$, then there exists a $\chi_{dd}$-coloring $\{V_1, V_2, \{x\}\}$ of $G$ such that $V_1$ contains all the leaves of $x$. Now, we get two vertices $u$ and $v$ on $C$, such that $u \in V_1, v \in V_2$, both $u$ and $v$ are not adjacent to $x$. Clearly, $v$ does not dominate any color class and the color class $V_1$ does not be dominated by any vertex, which is a contradiction. Thus, $G$ has no support vertices and $G = C$.  By Theorem \ref{th2}(2), $G = C_5$. So, the theorem follows. 
\hfill $\square$

\begin{figure}[h]
\centering\includegraphics[width=0.3\linewidth]{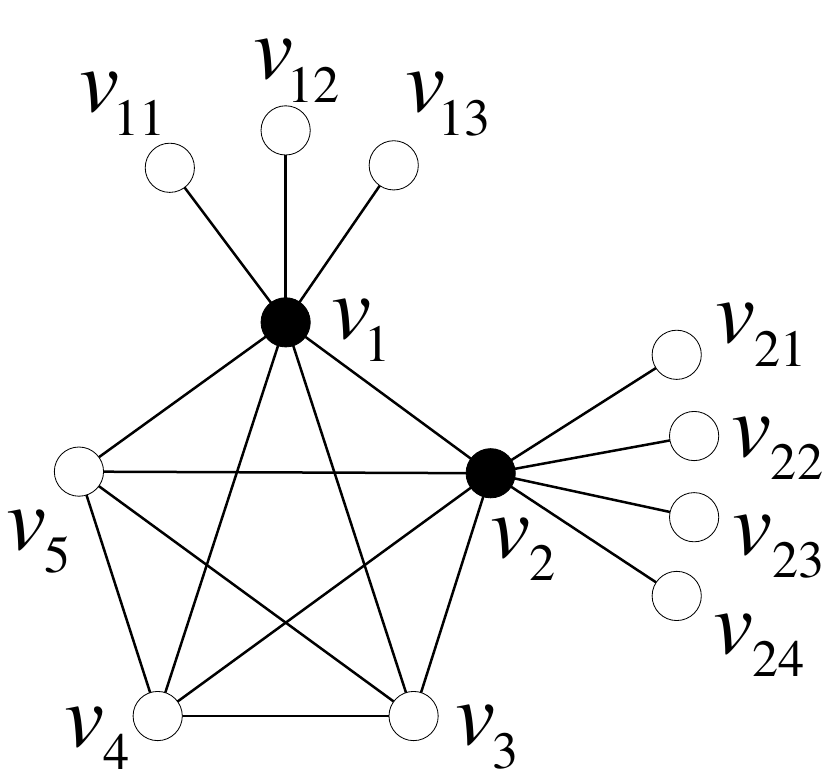}
\caption{A instance of $K_5^2$}
\label{f4}
\end{figure}

For the complete graph $K_n$, we know that $\chi_{dd}(K_n)=\chi(K_n)=n$. Next, we construct a family of graphs by attaching leaves at some vertices of the complete graph. We denote $\mathcal{K}_n^m$ the family of graphs obtained by attaching leaves at $m$ vertices of $K_n$, $1 \leq m \leq n$. We take no account of the number of leaves attached at any vertex in the notation, since it does not impact the domination chromatic number. Morever, we denote any element in $\mathcal{K}_n^m$ by $K_n^m$. For example, a instance of $K_5^2$ is shown in Figure \ref{f4}.

\begin{theorem}
\label{th8}
For $m \leq \lfloor\frac{n}{2}\rfloor$, $\chi_{dd}(K_n^m)=\chi(K_n^m)$.
\end{theorem}

Proof. For any $1 \leq m \leq n$, $K_n^m$ is $n$-colorable. So, $\chi(K_n^m) = n$. Next, we consider a domination coloring of $K_n^m$. On one hand, each vertex attached leaves should be partitioned into a singleton color class, since each of these leaves has to dominate a color class formed by its only neighbor.  On the other hand, leaves attached to different vertices have to be partitioned into different color classes, otherwise, there exist no vertex dominating the color class. Thus, at most $\lfloor\frac{n}{2}\rfloor$ vertices can be attached leaves of $K_n$, in order to guarantee that $K_n^m$ is $n$-domination colorable. The result follows.
\hfill $\square$



\section{Conclusion and Further Research}

In this paper, we propose the concept of domination coloring based on the dominator coloring and dominated coloring. Basic results and properties of the domination coloring number are studied. Also, we characterize two classes of graph that have the property that $\chi_{dd}(G) = \chi(G)$.

The following are some interesting problems for further investigation.

\begin{itemize}
\item For what graphs does $\chi_{dd}(G) = \chi(G)$?
\item For what graphs does $\chi_{dd}(G) = \gamma(G)$?
\item For what graphs does $\chi_{dd}(G) = \chi(G) \cdot \gamma(G)$?
\end{itemize}

\noindent\textbf{Acknowledgements}

The authors acknowledge support from Peking University. The authors also would like to thank referees for their useful suggestions. 





\bibliographystyle{model1-num-names}
\bibliography{sample}

\begin{thebibliography}{16}
\expandafter\ifx\csname natexlab\endcsname\relax\def\natexlab#1{#1}\fi
\providecommand{\bibinfo}[2]{#2}
\ifx\xfnm\relax \def\xfnm[#1]{\unskip,\space#1}\fi
\bibitem[{Bondy and Murty(2008)}]{Bondy:2008}
\bibinfo{author}{J.~A. Bondy}, \bibinfo{author}{U.~S.~R. Murty},
  \bibinfo{title}{Graph Theory}, \bibinfo{publisher}{Springer},
  \bibinfo{year}{2008}.
\bibitem[{Chen and Schelp(1998)}]{Chen:1998}
\bibinfo{author}{G.~Chen}, \bibinfo{author}{R.~H. Schelp},
  \bibinfo{title}{Vertex coloring with a distance restriction},
  \bibinfo{year}{1998}.
\bibitem[{Malaguti and Toth(2010)}]{Malaguti:2010}
\bibinfo{author}{E.~Malaguti}, \bibinfo{author}{P.~Toth},
\newblock \bibinfo{title}{A survey on vertex coloring problems},
\newblock \bibinfo{journal}{International Transactions in Operational Research}
  \bibinfo{volume}{17} (\bibinfo{year}{2010}) \bibinfo{pages}{1--34}.
\bibitem[{Haynes et~al.(1998{\natexlab{a}})Haynes, Hedetniemi, and
  Slater}]{Haynes:199801}
\bibinfo{author}{T.~W. Haynes}, \bibinfo{author}{S.~T. Hedetniemi},
  \bibinfo{author}{P.~J. Slater}, \bibinfo{title}{Fundamentals of Domination in
  Graphs}, \bibinfo{year}{1998}{\natexlab{a}}.
\bibitem[{Haynes et~al.(1998{\natexlab{b}})Haynes, Hedetniemi, and
  Slater}]{Haynes:199802}
\bibinfo{author}{T.~W. Haynes}, \bibinfo{author}{S.~T. Hedetniemi},
  \bibinfo{author}{P.~J. Slater}, \bibinfo{title}{Domination in Graphs: Volume
  2: Advanced Topics}, \bibinfo{year}{1998}{\natexlab{b}}.
\bibitem[{Chellali and Volkmann(2004)}]{Chellali:2004}
\bibinfo{author}{M.~Chellali}, \bibinfo{author}{L.~Volkmann},
\newblock \bibinfo{title}{Relations between the lower domination parameters and
  the chromatic number of a graph},
\newblock \bibinfo{journal}{Discrete Mathematics} \bibinfo{volume}{274}
  (\bibinfo{year}{2004}) \bibinfo{pages}{1--8}.
\bibitem[{Hedetniemi et~al.(2009)Hedetniemi, Hedetniemi, Laskar, Mcrae, and
  Wallis}]{Hedetniemi:2009}
\bibinfo{author}{S.~M. Hedetniemi}, \bibinfo{author}{S.~T. Hedetniemi},
  \bibinfo{author}{R.~Laskar}, \bibinfo{author}{A.~A. Mcrae},
  \bibinfo{author}{C.~K. Wallis},
\newblock \bibinfo{title}{Dominator partitions of graphs},
\newblock \bibinfo{journal}{Journal of Combinatorics, Information \& System
  Sciences} \bibinfo{volume}{34} (\bibinfo{year}{2009}).
\bibitem[{Gera et~al.(2006)Gera, Rasmussen, and Horton}]{Gera:2006}
\bibinfo{author}{R.~M. Gera}, \bibinfo{author}{C.~Rasmussen},
  \bibinfo{author}{S.~Horton},
\newblock \bibinfo{title}{Dominator colorings and safe clique partitions},
\newblock \bibinfo{journal}{Congressus Numerantium} \bibinfo{volume}{181}
  (\bibinfo{year}{2006}) \bibinfo{pages}{19--32}.
\bibitem[{Gera(2007{\natexlab{a}})}]{Gera:200701}
\bibinfo{author}{R.~M. Gera},
\newblock \bibinfo{title}{On dominator colorings in graphs},
\newblock \bibinfo{journal}{Graph Theory Notes N. Y.} \bibinfo{volume}{52}
  (\bibinfo{year}{2007}{\natexlab{a}}) \bibinfo{pages}{25--30}.
\bibitem[{Gera(2007{\natexlab{b}})}]{Gera:200702}
\bibinfo{author}{R.~M. Gera},
\newblock \bibinfo{title}{On the dominator colorings in bipartite graphs},
\newblock \bibinfo{journal}{International Conference on Information Technology}
   (\bibinfo{year}{2007}{\natexlab{b}}).
\bibitem[{Chellali and Maffray(2012)}]{Chellali:2012}
\bibinfo{author}{M.~Chellali}, \bibinfo{author}{F.~Maffray},
\newblock \bibinfo{title}{Dominator colorings in some classes of graphs},
\newblock \bibinfo{journal}{Graphs \& Combinatorics} \bibinfo{volume}{28}
  (\bibinfo{year}{2012}) \bibinfo{pages}{97--107}.
\bibitem[{Merouane and Chellali(2012)}]{Boumediene:2012}
\bibinfo{author}{H.~B. Merouane}, \bibinfo{author}{M.~Chellali},
\newblock \bibinfo{title}{On the dominator colorings in trees},
\newblock \bibinfo{journal}{Discussiones Mathematicae Graph Theory}
  \bibinfo{volume}{32} (\bibinfo{year}{2012}) \bibinfo{pages}{677--683}.
\bibitem[{Arumugam et~al.(2012)Arumugam, Bagga, and
  Chandrasekar}]{ARUMUGAM:2012}
\bibinfo{author}{S.~Arumugam}, \bibinfo{author}{J.~Bagga},
  \bibinfo{author}{K.~R. Chandrasekar},
\newblock \bibinfo{title}{On dominator colorings in graphs},
\newblock \bibinfo{journal}{Proceedings - Mathematical Sciences}
  \bibinfo{volume}{122} (\bibinfo{year}{2012}) \bibinfo{pages}{561--571}.
\bibitem[{Guillaume et~al.(2017)Guillaume, Houcine, Mohammed, and
  Hamamache}]{Bagan:2017}
\bibinfo{author}{B.~Guillaume}, \bibinfo{author}{B.~M. Houcine},
  \bibinfo{author}{H.~Mohammed}, \bibinfo{author}{K.~Hamamache},
\newblock \bibinfo{title}{On some domination colorings of graphs},
\newblock \bibinfo{journal}{Discrete Applied Mathematics}
  (\bibinfo{year}{2017}).
\bibitem[{Merouane et~al.(2015)Merouane, Haddad, Chellali, and
  Kheddouci}]{Boumediene:2015}
\bibinfo{author}{H.~B. Merouane}, \bibinfo{author}{M.~Haddad},
  \bibinfo{author}{M.~Chellali}, \bibinfo{author}{H.~Kheddouci},
\newblock \bibinfo{title}{Dominated colorings of graphs},
\newblock \bibinfo{journal}{Graphs \& Combinatorics} \bibinfo{volume}{3}
  (\bibinfo{year}{2015}) \bibinfo{pages}{713--727}.
\bibitem[{Appel and Haken(1976)}]{Appel:1976}
\bibinfo{author}{K.~Appel}, \bibinfo{author}{W.~Haken},
\newblock \bibinfo{title}{Every planar map is four colorable},
\newblock \bibinfo{journal}{Bulletin of the American Mathematical Society}
  \bibinfo{volume}{5} (\bibinfo{year}{1976}) \bibinfo{pages}{711--712}.

\end{thebibliography}







\end{document}